\documentclass[aps,prl,twocolumn,superscriptaddress]{revtex4-2}

\usepackage[utf8]{inputenc}
\usepackage{bbm}

\usepackage{placeins}
\usepackage{xcolor}
\usepackage{graphicx}
\usepackage{hyperref}
\usepackage[export]{adjustbox}

\usepackage{mathtools}
\usepackage{amsmath}
\usepackage{amssymb}
\usepackage{physics}
\usepackage{braket}

\hypersetup{pdfborder = {0 0 0}}
\hypersetup{colorlinks,citecolor=blue,filecolor=blue,linkcolor=blue,urlcolor=blue}

\begin{document}
\title{Many-body quantum non-Markovianity}

\author{Jonathan Brugger}
\affiliation{Physikalisches Institut, Albert-Ludwigs-Universität Freiburg,\\ Hermann-Herder-Straße 3, 79104 Freiburg, Germany}
\affiliation{EUCOR Centre for Quantum Science and Quantum Computing, Albert-Ludwigs-Universität Freiburg,\\ Hermann-Herder-Straße 3, 79104 Freiburg, Germany}

\author{Christoph Dittel}
\affiliation{Physikalisches Institut, Albert-Ludwigs-Universität Freiburg,\\ Hermann-Herder-Straße 3, 79104 Freiburg, Germany}
\affiliation{EUCOR Centre for Quantum Science and Quantum Computing, Albert-Ludwigs-Universität Freiburg,\\ Hermann-Herder-Straße 3, 79104 Freiburg, Germany}

\author{Andreas Buchleitner}
\affiliation{Physikalisches Institut, Albert-Ludwigs-Universität Freiburg,\\ Hermann-Herder-Straße 3, 79104 Freiburg, Germany}
\affiliation{EUCOR Centre for Quantum Science and Quantum Computing, Albert-Ludwigs-Universität Freiburg,\\ Hermann-Herder-Straße 3, 79104 Freiburg, Germany}

\begin{abstract}
	We port the concept of non-Markovian quantum dynamics to the many-particle realm, by a suitable decomposition of the many-particle Hilbert space.
	We show how the specific structure of many-particle states determines
	the observability of non-Markovianity by single- or many-particle observables, and discuss a realization in a readily implementable few-particle set-up.
\end{abstract}

\date{\today}

\maketitle

Non-Markovian behavior  \cite{NonMar3} is the partial restoration of an open quantum system's memory of its past. 
In general, we expect that an open quantum system -- widely (though not exclusively) understood 
as living on a small number of degrees of freedom, as opposed to the many degrees of freedom of an environment or bath it is coupled to -- tends to irreversibly loose its 
memory, since any information leaking to the environment will quickly disperse and not relocalize on the system degrees 
of freedom \cite{AtomPhotonInteractions,kolovsky1994,abu2003}. 
However, this intuition is reliable only if
the number of degrees of freedom associated with system and environment, as well as the associated spectral structures, are distinct, and when 
system and environment 
do not easily correlate. Consequently, non-Markovian behavior is naturally expected, e.g., in large molecular structures 
\cite{rebentrost2009,walschaers2013,hongbinchen2015,levi2015,roden2016},
where different degrees of freedom are strongly coupled and typically non-separable, with the consequence that Markovian 
master equation-like descriptions 
(successfully employed 
for many quantum optical applications \cite{AtomPhotonInteractions}) turn unreliable. Given the ever improving experimental resolution of the dynamics
of diverse 
multi-component quantum systems \cite{rozzi2013,meinert2014,maly2016,wittemer2018,bruder2019,wittmann2020}, there is a strong incentive to improve our understanding of 
non-Markovianity, and to identify observables which allow for an unambiguous identification 
of non-Markovian effects.

While the prerequisites for Markovian behavior have been known for long \cite{AtomPhotonInteractions,kolovsky1994}, the concept of non-Markovianity needed to be formalized, and many of its subtleties have been clarified \cite{NonMar5, NonMar6, NonMar7, NonMar1, NonMar2, NonMar8, NonMar3, NonMar4} during recent years.
Yet, the specific impact of the generic structural features of many-particle systems upon the manifestations of non-Markovianity hitherto remained unexplored. 
Our present contribution
specifically addresses non-Markovianity in the many-particle context.

We 
rely on the 
quantification of non-Markovian behavior in terms of information back-flow from the environment into the system, through the time dependence of the trace distance \cite{NonMar6,NonMar3}
\begin{gather}
	\begin{gathered}
		D(\rho, \sigma) = \frac{1}{2} \tr |\rho - \sigma|
	\end{gathered}\label{eq1}
\end{gather}
of initially distinct system states $\rho$ and $\sigma$, with $|M| = \sqrt{MM^\dagger}$ the positive square root of a positive semi-definite operator. 
$D$ is a metric on the space of density matrices, with 
$D(\rho, \sigma) = 0$ if and only if $\rho = \sigma$, and $D(\rho, \sigma) = 1$ if and only if $\rho$ and $\sigma$ have orthogonal support \cite{NonMar3}. As 
an exhaustive measure for the distinguishability of two quantum states -- by any type of measurement -- it quantifies their 
distinctive information content. However, when state tomography turns unaffordable, e.g. due to the underlying Hilbert space dimension \cite{haeffner2005}, 
it is a priori often unclear which observable can expose such distinctive information most efficiently, and particularly so when dealing with many-particle states. 

Intuitively, though, information flow is easily associated with the exchange of excitations in different degrees of freedom, and we will here build on this intuition by 
considering the specific -- and experimentally feasible \cite{preiss2019} -- example of few fermions or bosons loaded into 
the asymmetric, one-dimensional double-well potential depicted in Fig.~\ref{fig1}(a):
\begin{figure}[t]
	\centering
	\includegraphics[width=\linewidth]{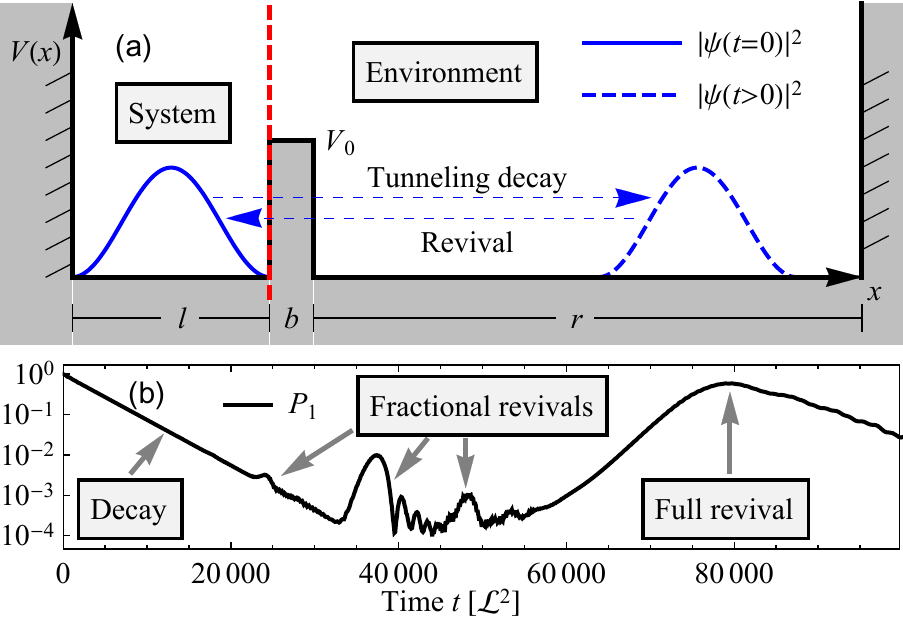}
	\caption{Tunneling 
	in a one-dimensional asymmetric double-well potential.
	(a) Sketch: A particle prepared (continuous blue line, $|\psi(t=0)|^2$) in the left well (system) tunnels into the right well (environment), is reflected at the right boundary, and possibly re-enters the system (dashed blue line, $|\psi(t>0)|^2$). 
	(b) Semilogarithmic plot of the particle's probability $P_1$ to be found in the system,
	if initially prepared in the isolated system's ground state $\ket{1}$,
	for 
	$l = 50 \, \mathcal{L}$, $b = 2 \, \mathcal{L}$, $r = 4000 \, \mathcal{L}$ and $V_0 = 0.1 \, \mathcal{L}^{-2}$.
	}\label{fig1}
\end{figure}
Left and right potential wells of width $l$ and $r$ each define local mode structures which we associate with the system's and environmental degrees of freedom, respectively. They are separated by a finite rectangular barrier of width $b$ and height $V_0$, which is part of the environment. As shown earlier \cite{Potential0, StefanHunn2, StefanHunn1}, this model allows an exact spectral treatment of the decay dynamics of an open many-body system with a continuously tunable, discrete to quasi-continuous spectral structure of the environment. System-environment coupling is determined by $b$ and $V_0$, and weak coupling and a quasi-continuous spectrum of the environment, in the limit of large $r$, justify the intuition of the system being opened to couple to the environment's degree of freedom. 

Single-particle dynamics are obtained
from exact numerical diagonalization \cite{LongPaper} of the single-particle Hamiltonian $H_\mathrm{sp}(x) = - \partial^2/\partial x^2 + V(x)$, after discretization in a suitable finite element basis \cite{BartelsFEM, FEM}. 
While we will elaborate elsewhere \cite{LongPaper} on how to control the arising 
non-Markovian tunneling dynamics by tuning, through variable $r$, the transition to a quasi-continuum, we here focus on a fixed, finite width $r$, which generates the typical behavior. In all subsequent simulations we choose the parameters $l = 50 \, \mathcal{L}$, $b = 2 \, \mathcal{L}$, $r = 4000 \, \mathcal{L}$ and $V_0 = 0.1 \, \mathcal{L}^{-2}$ (in natural units, 
with $\hbar \equiv 1$ and mass $m \equiv 1/2$, and in terms of the characteristic experimental length scale $\mathcal{L}$ \cite{StefanHunn2}), which, in particular, establish the quasi-continuous limit for the environment's spectrum (notwithstanding the hard-wall boundary condition of the right well's outer 
confinement, which clearly induces non-Markovian behavior on sufficiently long time scales \cite{StefanHunn1,StefanHunn2}).
The 
dynamics can then be conceived in the Hilbert space $\mathcal{H} = \mathcal{H}_\mathrm{S} \oplus \mathcal{H}_\mathrm{E}$, with $\mathcal{H}_\mathrm{S(E)}$ representing the system (environment). Since
$\mathcal{H}$ is given by a direct sum, 
any pure state of 
system and environment can be written as $\ket{\psi} = c_1 \Pi_\mathrm{S} \ket{\psi} + c_0 \Pi_\mathrm{E} \ket{\psi}$, with $\Pi_\mathrm{S(E)}$ the projector onto $\mathcal{H}_\mathrm{S(E)}$, and the probability $P_1 = |c_1|^2$ to find the particle in 
the 
system. 

\begin{figure*}[t]
	\centering
	\includegraphics[width=\linewidth]{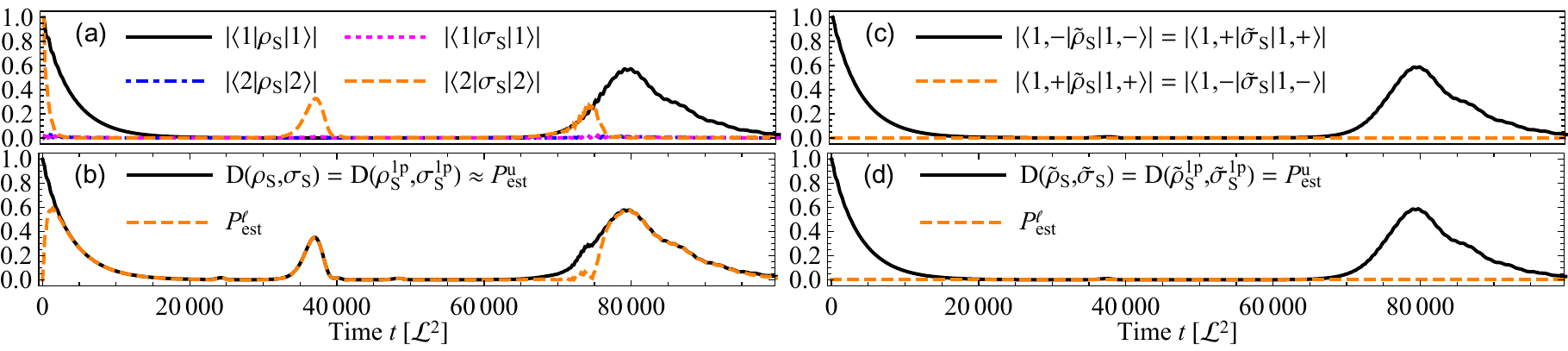}
	\caption{Non-Markovian single-particle tunneling in an asymmetric double-well potential (see Fig.~\ref{fig1}). The revivals (a) of the auto- 
	($|\langle 1|\rho_{\rm S}|1\rangle |$,  $|\langle 2|\sigma_{\rm S}|2\rangle |$) and cross correlation functions ($|\langle 2|\rho_{\rm S}|2\rangle |$,  
	$|\langle 1|\sigma_{\rm S}|1\rangle |$) of a single particle initially prepared
	in the ground state $\ket{1}$ or in the first excited state $\ket{2}$, respectively, of the isolated left potential well and reduced to the system 
	(i.e., the left well's)
	degrees of freedom are faithfully reflected by (b) the trace distance $D(\rho_\mathrm{S},\sigma_\mathrm{S})$, Eq.~(\ref{eq1}), as well as, almost everywhere, by its estimators 
	(\ref{eq6},\ref{eq7},\ref{eq9}) ($P^\mathrm{u}_\mathrm{est} \approx D(\rho_\mathrm{S},\sigma_\mathrm{S})$ \cite{LongPaper}, with deviations smaller than $10^{-2}$). 
	Only at $t \approx 74\,000 \,\mathcal{L}^2$ is the lower bound $P^\mathrm{\ell}_\mathrm{est}$, Eq.~(\ref{eq6}), not tight, since it does not resolve single state
	population differences, but only those of the total particle number probabilities in the left well. For the same reason is $P^\mathrm{\ell}_\mathrm{est}$ unable to discriminate 
	single particle states launched in the ground state $\ket{1}$, but labeled with mutually orthogonal states $\ket{\pm}$ in an additional degree of freedom: (d) Only the estimators 
	(\ref{eq7},\ref{eq9}) and the trace distance $D(\tilde{\rho}_\mathrm{S},\tilde{\sigma}_\mathrm{S})$ sense the revivals featured by (c) the 
	autocorrelation functions $|\langle 1,-|\tilde{\rho}_{\rm S}|1,-\rangle |$ and $|\langle 1,+|\tilde{\sigma}_{\rm S}|,1+\rangle |$.
	}\label{fig2}
\end{figure*}

Figure \ref{fig1}(b) shows the time evolution of $P_1$ after initializing the particle in the ground state $\ket{1}$ of the isolated system. We observe an exponential decay, followed by low-amplitude (note the semi-logarithmic scale) partial (fractional) \cite{FractionalRevival2}
revivals, and, subsequently, a pronounced full revival.
Fractional and full revivals are due to the coherent superposition of single particle amplitudes reflected at the barrier and at the right boundary of the environment, 
with an admixture of non-vanishing excited state amplitudes of the system degree of freedom. The latter 
define the fractional revival times
in Fig.~\ref{fig1}(b), and are individually enhanced when launching the dynamics in an excited system state, as, e.g., $\ket{2}$ in  Fig.~\ref{fig2}(a).
Since such revivals express excitation and, thus, information backflow into the system degree of freedom, they are indicative of non-Markovianity, as we quantify further down.

Let us, however, first turn towards the general many-particle problem, and, in particular, to the intriguing, non-trivial structure
of the many-particle state space: The many-particle dynamics of any number $N$ of identical bosons (fermions) play in the Fock space $\Gamma^\mathrm{b(f)}(\mathcal{H})$ \cite{AQMSchwabl}
constructed from the single-particle Hilbert space $\mathcal{H} = \mathcal{H}_\mathrm{S} \oplus \mathcal{H}_\mathrm{E}$, and can be factorized \cite{Mattia1,Mattia2} as
\begin{gather}
	\Gamma^\mathrm{b(f)}(\mathcal{H}_\mathrm{S} \oplus \mathcal{H}_\mathrm{E}) \cong \Gamma^\mathrm{b(f)}(\mathcal{H}_\mathrm{S}) \otimes \Gamma^\mathrm{b(f)}(\mathcal{H}_\mathrm{E}). \label{eq2}
\end{gather}
This \emph{tensor product} structure enables the interpretation as an open quantum system, despite the \emph{direct sum} structure of $\mathcal{H}$ 
(rather reminiscent of an atomic ionization problem \cite{abu1995}).
Unitary dynamics and a well-defined particle number on the combined system and environment degrees of freedom restrict the time evolution to the effective $N$-particle space
\begin{gather}
	\mathcal{H}_\mathrm{eff}^N = \bigoplus_{k = 0}^N \left(\mathcal{H}_\mathrm{S}^{\otimes k} \otimes \mathcal{H}_\mathrm{E}^{\otimes(N-k)}\right)_\mathrm{s(a)} \subsetneq \Gamma^\mathrm{b(f)}(\mathcal{H}),\label{eq3}
\end{gather}
with the (anti-)symmetrization ${\mathcal X}_\mathrm{s(a)}$ of a Hilbert space $\mathcal X$. As a consequence, every pure state of system 
and environment can be written as $\ket{\psi} = \sum_{k = 0}^N c_k \ket{\psi_k} \in \mathcal{H}_\mathrm{eff}^N$, with $\ket{\psi_k}$ a state with $k$ out of $N$ particles 
confined to the system.
To assess the non-Markovian character of the system dynamics, we need to trace $\ketbra{\psi}{\psi}$ over the environment, to produce the reduced 
system state
\begin{gather}
	\begin{gathered}
		\rho_\mathrm{S} = \mathrm{tr}_\mathrm{E} \ketbra{\psi}{\psi} = \sum_{k=0}^N |c_k|^2 \, \rho_k,
	\end{gathered} \label{eq4}
\end{gather}
with $\rho_k = \mathrm{tr}_\mathrm{E} \ketbra{\psi_k}{\psi_k}$.
The partial trace $\mathrm{tr}_\mathrm{E}$ is natural for the tensor product in Eq.~\eqref{eq2}, and well-defined for $\ketbra{\psi}{\psi}$ by restriction to $\mathcal{H}_\mathrm{eff}^N$. Intuitively, it can be thought of as tracing over the $N - k$ particles in the environment for each summand in Eq.~\eqref{eq3}.
$\rho_\mathrm{S}$ exhibits block-diagonal structure, since states of the environment corresponding to different particle numbers are orthogonal \cite{LongPaper}.

The trace distance $D( \rho_\mathrm{S}, \sigma_\mathrm{S})$ between two reduced states $\rho_\mathrm{S}= \sum_k |c_k|^2 \rho_k$ and $\sigma_\mathrm{S}= \sum_k |d_k|^2 \sigma_k$ of the system can now be evaluated for arbitrary particle number, particle type (bosons or fermions), degree of (in)distinguishability, and interaction strength and type, as 
$D( \rho_\mathrm{S}, \sigma_\mathrm{S}) = \sum_{k = 0}^N D(|c_k|^2 \rho_k, |d_k|^2 \sigma_k)$. 
Whenever $D( \rho_\mathrm{S}, \sigma_\mathrm{S})$ increases as a function of time, this signals non-Markovian behavior.
Although the block-diagonal structure of $\rho_\mathrm{S}$ and $\sigma_\mathrm{S}$ significantly reduces the computational complexity of the trace distance $D(\rho_\mathrm{S}, \sigma_\mathrm{S})$, it remains non-trivial to evaluate, especially for the dynamics of many \emph{interacting} particles \cite{LongPaper}. However, as we show in the following, this burden can often be alleviated by relating the trace distance $D(\rho_\mathrm{S}, \sigma_\mathrm{S})$ to computationally and experimentally more readily accessible quantities.

From Eq.~\eqref{eq4} we obtain the probabilities $P_k\left( \rho_\mathrm{S} \right) = |c_k|^2$ to find \emph{exactly} $k$ out of $N$ particles in the system, which constitute simple many-particle observables (given 
number-resolving detectors), and 
are useful to distinguish Markovian from non-Markovian many-body dynamics \cite{Note1}. A straightforward calculation \cite{LongPaper} shows that the trace distance is bounded by
\begin{gather}
	P_\mathrm{est}^\mathrm{\ell} \leq D( \rho_\mathrm{S}, \sigma_\mathrm{S} ) \leq P_\mathrm{est}^\mathrm{u},\label{eq5}
\end{gather}
with the lower and upper bounds
given by
\begin{gather}
	P_\mathrm{est}^\mathrm{\ell} = \sum_{k=0}^N \frac{\left| |c_k|^2 - |d_k|^2 \right|}{2}\label{eq6}
\end{gather}
and
\begin{gather}
	P_\mathrm{est}^\mathrm{u} = 1 - \frac{|c_0|^2 + |d_0|^2}{2} + \frac{\left| |c_0|^2 - |d_0|^2 \right|}{2},\label{eq7}
\end{gather}
respectively. Intuitively, $P_\mathrm{est}^\mathrm{\ell}$ is the sum of \emph{minimal} trace distances within the blocks in \eqref{eq4}, given each by the associated
population differences. Analogously, $P_\mathrm{est}^\mathrm{u}$ is the sum of the \emph{exact} trace distance in the one-dimensional block $k = 0$, and of
the \emph{maximal} trace distances within all blocks $k \geq 1$, again given by population differences.

Similarly, we can bound the trace distance $D(\rho_\mathrm{S}, \sigma_\mathrm{S})$ in terms of single particle observables. To this end we consider the \emph{reduced single-particle density matrix} (RSPDM)
\begin{gather}
	\begin{gathered}
		\rho^\mathrm{1p}_\mathrm{S} =
		|c_0|^2 \, \rho_0 + |c_1|^2 \, \rho_1 + \sum_{k = 2}^N |c_k|^2 \,  \tr_{2,\dots,k}\rho_k
	\end{gathered}\label{eq8}
\end{gather}
obtained from the system's state~\eqref{eq4} by tracing out all but one particle \cite{RSPDM1,RSPDM2}. It describes a potentially mixed state with up to one particle in the system
 and offers a natural way to compare states on a single-particle level, since the expectation value (with respect to $\rho_\mathrm{S}$) of any single-particle observable (like, e.g., the projection $\ketbra{1}{1}$ onto the single-particle ground state) can be 
 inferred from it \cite{RSPDM2}. Using the contractivity of $D$ under trace-preserving quantum operations \cite{TraceDistanceContraction, NielsenChuang} we find
\begin{gather}
	D(\rho^\mathrm{1p}_\mathrm{S}, \sigma^\mathrm{1p}_\mathrm{S}) \leq D(\rho_\mathrm{S}, \sigma_\mathrm{S}),\label{eq9}
\end{gather}
again bounding the trace distance from below.

\begin{figure}[t]
	\centering
	\includegraphics[width=\linewidth]{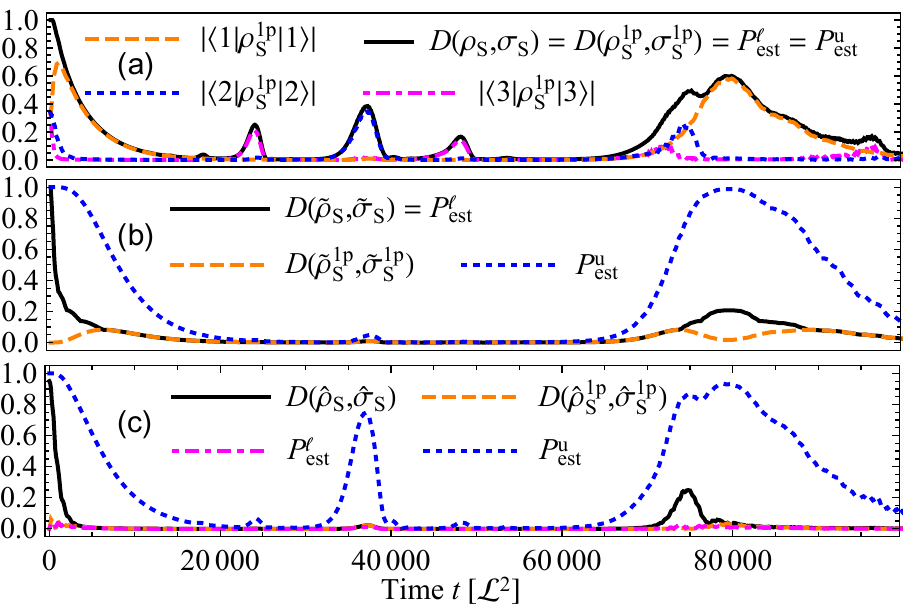}
	\caption{Non-Markovian many-particle tunneling: (a) Three indistinguishable fermions launched in the (fermionic) ground state of the isolated left well (see Fig.~\ref{fig1}) 
	exhibit clear revivals of the reduced single particle autocorrelation functions $|\langle j|\rho_{\rm S}^{\rm 1p}|j\rangle |$, $j=1,2,3$,  
	of the left well's ground, first and second excited single particle states. This gives rise to non-Markovian behavior as clearly manifest in the state's trace distance
	$D(\rho_\mathrm{S}, \sigma_\mathrm{S})$
	from the left well's (time-invariant) many-particle vacuum state $\sigma_\mathrm{S} = \ketbra{0_\mathrm{S}}{0_\mathrm{S}}$.
	The trace distance estimators $D(\tilde{\rho}^{1p}_\mathrm{S}, \tilde{\sigma}^{1p}_\mathrm{S})$, $P_\mathrm{est}^\mathrm{\ell}$, 
	$P_\mathrm{est}^\mathrm{u}$, (\ref{eq6},\ref{eq7},\ref{eq9}), are tight, since $d_k = \delta_{0k}$ for the many-particle vacuum state \cite{LongPaper}.
	(b) and (c) monitor the trace distance and its estimators (\ref{eq6},\ref{eq7},\ref{eq9}) for two pairs of bosonic many-particle states: The trace distance 
	$D(\tilde{\rho}^{1p}_\mathrm{S}, \tilde{\sigma}^{1p}_\mathrm{S})$ of 
	the reduced single particle states of the four and five particle states launched in the bosonic ground states (b) of the isolated left well, 
	$\tilde{\rho}(0) = \tilde{\rho}_\mathrm{S}(0) = \ketbra{1}{1}^{\otimes 4}$ and $\tilde{\sigma}(0) = \tilde{\sigma}_\mathrm{S}(0) = \ketbra{1}{1}^{\otimes 5}$, respectively,
	only barely detects the many-body revival in the system ground state. To detect non-Markovianity by comparison of (c) the many-particle dynamics of six particles 
	prepared in 
	$\hat{\rho}(0) = \hat{\rho}_\mathrm{S}(0) = \mathcal{S}(\ketbra{1}{1}^{\otimes 3} \otimes \ketbra{2}{2}^{\otimes 3})$ and 
	$\hat{\sigma}(0) = \hat{\sigma}_\mathrm{S}(0) = \ketbra{1}{1}^{\otimes 3} \otimes \ketbra{2}{2}^{\otimes 3}$, respectively, which differ only by their (un-) symmetrized character 
	($\mathcal S$ the bosonic symmetrization operator), bona fide many-particle observables need to be interrogated.
}\label{fig3}
\end{figure}

Equations (\ref{eq5}--\ref{eq9}) thus provide bounds on $D(\rho_\mathrm{S}, \sigma_\mathrm{S})$, and, consequently, on the non-Markovianity of the system dynamics \footnote{The experimentally feasible bounds (\ref{eq5}-\ref{eq9}) are designed to witness non-Markovian behavior: any two points in time $t_0 < t_1$ for which either $P_\mathrm{est}^\mathrm{u}(t_0) < P_\mathrm{est}^\mathrm{\ell}(t_1)$ or $P_\mathrm{est}^\mathrm{u}(t_0) < D[\rho_\mathrm{S}(t_1), \sigma_\mathrm{S}(t_1)]$ imply a temporally increasing trace distance $D(\rho_\mathrm{S}, \sigma_\mathrm{S})$, and therefore -- by definition -- non-Markovian behavior.}, which can be assessed by monitoring simple features of the counting statistics or of single particle observables like the ground state population of the system. The tightness of these bounds is inspected for different single- and many-particle states, in Fig.~\ref{fig2} and Fig.~\ref{fig3}, respectively. To consider (partially) (in)distinguishable particles \cite{WaveParticleDuality}, we further equip the particles with an internal degree of freedom, e.g. $\mathcal{H}_\mathrm{int} = \mathrm{span}\{\ket{+}, \ket{-}\}$, not coupled to their external degree of freedom (i.e., $H_\mathrm{sp}(x)$ is independent of the particle's internal state). 
While all stated conceptual observations and analytical results apply, in particular, also for interacting particles, we restrict our subsequent 
numerical examples to the non-interacting case, such that 
many-particle eigenstates are (anti-)symmetrized product states of single-particle eigenstates 
(interacting particles will be considered elsewhere \cite{LongPaper}). 

Starting with single-particle dynamics, Fig.~\ref{fig2} (a,b) compares the evolution of the auto- and cross correlation functions of two pure single particle states, launched in the 
system's ground and first excited states, respectively, upon trace over the environment, to the time evolution of their trace distance. We see that the revival dynamics of the system 
state populations in (a) is faithfully reflected by the trace distance in (b), and almost everywhere reproduced by the lower bound $P_\mathrm{est}^\mathrm{\ell}$, thus comforting our 
intuition that information backflow is synonymous to excitation backflow. However, we also see from the mismatch between $P_\mathrm{est}^\mathrm{\ell}$ and 
$D(\rho_\mathrm{S}, \sigma_\mathrm{S})$ at $t\simeq 74\,000 \,\mathcal{L}^2$, where both autocorrelation functions revive simultaneously, 
that $P_\mathrm{est}^\mathrm{\ell}$, which only monitors the population difference in the system, without resolving individual system state populations, 
is then too coarse grained a quantifier to distinguish both states. Likewise, the reviving trace distance of two pure single particle states both launched in the system's ground state, 
but labeled by mutually orthogonal states of an additional degree of freedom, is faithfully reproduced  
by $P_\mathrm{est}^\mathrm{u}$ and $D(\rho^\mathrm{1p}_\mathrm{S}, \sigma^\mathrm{1p}_\mathrm{S})$, 
while $P_\mathrm{est}^\mathrm{\ell}$ is blind for
this distinction, by its very construction. The latter is in contrast to the estimate $P_\mathrm{est}^\mathrm{u}$, which is (approximately) tight in Fig.~\ref{fig2}(d) (Fig.~\ref{fig2}(b)), as the particles are prepared in and return to (essentially) orthogonal single-particle modes of the system, thus (almost) realizing the maximal trace distance assumed in the derivation of~\eqref{eq7}.

On the many-particle level, a non-vanishing trace distance can have different physical causes,
including -- in our present case of non-interacting particles -- different constituting single-particle states, particle numbers, and symmetry properties. Let us 
inspect how to sense such differences using the bounds derived above.
Fig.~\ref{fig3}(a)
monitors the trace distance $D(\rho_\mathrm{S}, \ketbra{0_\mathrm{S}}{0_\mathrm{S}})$ of the reduced system state of
three non-interacting, 
indistinguishable fermions, launched in the system ground state, from the (time-invariant) many-particle system vacuum 
$\ket{0_{\rm S}}$, as well as the trace distance estimates (\ref{eq6},\ref{eq7},\ref{eq9}), which here all coincide \cite{LongPaper}.
Recombination(s) of the three-particle state into the ground, first, and second excited system state, as clearly reflected by the revivals of the reduced state single particle 
autocorrelation functions $| \langle j|\rho^{\rm 1p}_{\rm S}|j\rangle |$, induces non-Markovianity as unambiguously indicated by synchronous 
revivals of the above distance measures.

Fig.~\ref{fig3} (b) shows the time evolution of the trace distance of 
bosonic four- and five-particle states launched in the system ground states, $\tilde{\rho}(0) = \tilde{\rho}_\mathrm{S}(0) = \ketbra{1}{1}^{\otimes 4}$ 
and $\tilde{\sigma}(0) = \tilde{\sigma}_\mathrm{S}(0) = \ketbra{1}{1}^{\otimes 5}$, respectively. Non-Markovianity here stems from a many-particle repopulation 
of the system ground state, which, due to dispersion within the environment, stretches over a longer time interval, thus leading to an only mild revival of the 
states' trace distance. The reduced single particle states are barely discriminated by the single particle trace distance, on the rising and on the falling edge of the
repopulation of the system ground state, and cannot be distinguished at the maximum of the many-particle revival (by construction).

While the bosonic many-particle states in panel (b) are distinguished by their particle numbers, Fig.~\ref{fig3}(c) considers two $(N=6)$-particle
states of identical bosons \cite{WaveParticleDuality}, with identical and pairwise orthogonal internal states, respectively.
These states only differ if at least two particles in different single-particle states populate the system, 
and we therefore specifically consider 
$\hat{\rho}(0) = \hat{\rho}_\mathrm{S}(0) = \mathcal{S}(\ketbra{1}{1}^{\otimes 3} \otimes \ketbra{2}{2}^{\otimes 3})$ and 
$\hat{\sigma}(0) = \hat{\sigma}_\mathrm{S}(0) = \ketbra{1}{1}^{\otimes 3} \otimes \ketbra{2}{2}^{\otimes 3}$, with initially three particles in each, the 
ground and the first excited system state, respectively, and $\mathcal{S}$ the bosonic symmetrization operator.
Both states can be told apart only from their symmetry properties, by (reduced) $k$-particle trace distances $D(\hat{\rho}^{k\mathrm{p}}_\mathrm{S}, \hat{\sigma}^{k\mathrm{p}}_\mathrm{S})$ with $N\geq k > 1$ \footnote{For these initial states, we have reduced trace distances $D[\hat{\rho}^{k\mathrm{p}}_\mathrm{S}(0), \hat{\sigma}^{k\mathrm{p}}_\mathrm{S}(0)] \in \left\{ 0, \frac{3}{10}, \frac{3}{5}, \frac{4}{5}, \frac{9}{10} \right\}$
for $1 \leq k \leq 5$, 
and a full trace distance $D(\hat{\rho}_\mathrm{S}, \hat{\sigma}_\mathrm{S})=\frac{19}{20}$.
The latter value strictly smaller than one in particular shows that the two states here under scrutiny are not entirely orthogonal.}, and 
neither from single particle nor from number state populations. Consequently, none of the 
estimates (\ref{eq6},\ref{eq7},\ref{eq9}) provides a tight approximation of the states' actual trace distance $D(\hat{\rho}_\mathrm{S}, \hat{\sigma}_\mathrm{S})$,
which exhibits a weak revival at $t \approx 74\,000 \, \mathcal{L}^2$ due to the partially overlapping return of particles into the system's ground and first excited state (see also Fig.~\ref{fig2}(a)). Finally, the results in Fig.~\ref{fig3}(b,c) emphasize that $P_\mathrm{est}^\mathrm{u}$ -- by construction -- only guarantees a tight upper bound for $D(\rho_\mathrm{S}, \sigma_\mathrm{S})$ if either $\rho_\mathrm{S}$ or $\sigma_\mathrm{S}$ is sufficiently close to the many-particle vacuum.

We thus have seen that the 
non-Markovianity of $N$-body open system quantum evolution 
may be certifiable through experimentally readily accessible single particle observables like state populations or occupation numbers -- which 
can be inferred from the compared many-body states' reduced single-particle density matrices, but, depending on the 
choice of reference states, may also require to assess information inscribed into the latter's reduced $(k>1)$-particle system states, to discriminate on the level of  
many-body correlations and quantum statistical features. This hierarchical nesting of distinctive properties is resolved by the here derived many-body version 
of the reference states' trace distance -- given as a sum over trace distances of $k$-particle states, since the particle number is generally {\em not} conserved in an open 
many-body system. 
The decomposition into $k$-particle contributions, with $0\leq k\leq N$, crucially relies on our identification of the relevant
tensor structure of the many-body Fock space erected upon the single particle Hilbert space (itself given as a direct sum),
since it is this structure which reveals a many-body state's open system dynamics.



\vspace{20pt}

The authors thank Heinz-Peter Breuer and Moritz Richter for fruitful discussions. J. B. thanks the Studienstiftung des deutschen Volkes for support. C. D. acknowledges the Georg H. Endress foundation for financial support.

\bibliography{bibliography}

\end{document}